# Analyzing and modelling the AS-level Internet topology

**Shi Zhou & Raul J. Mondragon**
Department of Electronic Engineering
Queen Mary, University of London
Mile End Road, London, E1 4NS, United Kingdom
Email: {shi.zhou, r.j.mondragon}@elec.qmul.ac.uk
Tel: +44-2078825334, Fax: +44-2078827997

*Abstract* — Recently we introduced the rich-club phenomenon as a quantitative metric to characterize the tier structure of the Autonomous Systems level Internet topology (AS graph) and we proposed the Interactive Growth (IG) model, which closely matches the degree distribution and hierarchical structure of the AS graph and compares favourble with other available Internet power-law topology generators. Our research was based on the widely used BGP AS graph obtained from the Oregon BGP routing tables. Researchers argue that Traceroute AS graph, extracted from the traceroute data collected by the CAIDA's active probing tool, Skitter, is more complete and reliable. To be prudent, in this paper we analyze and compare topological structures of Traceroute AS graph and BGP AS graph. Also we compare with two synthetic Internet topologies generated by the IG model and the well-known Barabási-Albert (BA) model. Result shows that both AS graphs show the rich-club phenomenon and have similar tier structures, which are closely matched by the IG model, however the BA model does not show the rich-club phenomenon at all.

## I. INTRODUCTION

Analyzing and modelling the Internet topology is of immediate practical interest, since knowledge of the network's topological properties enables researchers to optimize network applications and to conduct more representative network simulations.

In the past few years, a lot of effort has been devoted to the characterization and modelling the Autonomous Systems (ASes) level Internet topology. Faloutsos *et al.* [1] discovered that the AS-level Internet topology has a power-law degree distribution, $P(k) \sim k^{-2.22}$, where node degree $k$ is the number of links a node has. Subramanian *et al.* [2], using a heuristic argument based on the commercial relationship between ASes, showed that the Internet has a tier structure. For example, Tier 1 consists of a 'core' of ASes that are densely connected to each other.

Researchers have introduced a number of Internet topology generators. Some of the generators [3] [4] [5] [6] are based on local properties (such as the degree distribution) and

This research is supported by the U.K. Engineering and Physical Sciences Research Council (EPSRC) under grant no. GR-R30136-01. The Data Kit #0204 used in this research was collected as part of CAIDA's Skitter initiative, http://www.caida.org. Support for skitter is provided by DARPA, NSF, and CAIDA membership.

others [7] [8] are based on large-scale structures (such as the tier structure). Tangmunarunkit *et al.* [9] found that structure-based network generators have a deliberately hierarchical structure, but their degree distributions are not power-laws. Whereas degree-based generators more accurately capture the large-scale structure of measured topologies. Their results were based on qualitative metrics and they recognized that there was a need for further studies to characterize network topologies.

Recently we reported the rich-club phenomenon [10] [11] as a quantitative metric to characterize the tier structure of power-law topologies without making any heuristic assumption on the interaction between network elements. Our study on the rich-club phenomenon revealed that degree-based models could generate networks with significantly different topological structures. Inspired by this result, we proposed the Interactive Growth (IG) model [12] [13]. This simple and dynamic model compares favourable with other available degree-based generators, because it matches both the power-law degree distribution and the hierarchical structure of the AS-level Internet topology.

There are currently two primary AS-level Internet connectivity graphs [19], which are BGP AS graph [14] [15] [16] and Traceroute AS graph [17] [18] [20] [22]. They use different methodologies for inferring network connectivity information. BGP AS graph is created from Border Gateway Protocol (BGP) inter-domain routing tables collected by University of Oregon [14]. Traceroute AS graph is extracted from the traceroute data collected by the CAIDA's active probing tool, Skitter [18].

Our research on the rich-club phenomenon was based on the widely used BGP AS graph. Researchers argue [19] [20] that BGP AS graph has limitations and Traceroute AS graph is more complete and reliable. To be prudent, in this paper we analyze and compare topological structures of traceroute AS graph and BGP AS graph. Also we compare with two synthetic Internet topologies generated by the well-known Barabási-Albert (BA) model [3] and the IG model. Result shows that both AS graphs show the rich-club phenomenon and have similar tier structures. The BA model does not have a densely interconnected core tier and therefore does not show the rich-club phenomenon at all. Whereas the IG model closely reproduces the tier structure of the AS-level Internet topology.

## II. BACKGROUND

Connections between participants in Internet communications can be abstracted in the dimension of network administration, which groups IP addresses into subnets, subnets into network prefixes and prefixes into autonomous systems (ASes). An autonomous system is the term in Border Gateway Protocol (BGP) for an entity that manages one or more networks and has a coherent policy for routing IP traffic both internally and to other ASes.

There are two types of data that predominate in research on AS-level Internet connectivity [19], which are BGP AS graph and Traceroute AS graph. The two graphs use different methodologies for inferring network connectivity information.

*A. BGP AS graph*

BGP AS graph are constructed from Internet interdomain BGP routing tables, which contain the information of connections from an AS to its immediate neighbours. The widely used BGP data are from the Oregon route server (*route-views.oregon-ix.net*) [14], which connects to several operational routers within the Internet for the purpose of collecting BGP routing tables.

BGP AS graph provided by the University of Oregon's Route Views Project [16] has been used in a number of studies on the Internet topology such as Faloutsos *et al*'s discovery [1] of the Internet power-law degree distribution and Barabási *et al*'s research [21] on the error and attack tolerance of the Internet.

Chen *et al.* [15] provided the so-called *extended map* [15] of BGP AS graph by using additional data sources, such as the Internet Routing Registry (IRR) data and the Looking Glass (LG) data. The IRR maintains individual ISP's (Internet Service Provider) routing information in several public repositories to coordinate global routing policy. The LG sites are maintained by individual ISPs to help troubleshoot Internet-wide routing problems. The extended map typically has 20-50% more links than the original BGP AS graph and provides a more complete picture of the Internet topology [15] [11].

BGP tables have the advantage that they are relatively easy to parse, process and comprehend. However, despite widespread public availability, BGP data has several limitations [19] [20] [22]. BGP tables do not reflect how traffic *actually* travels in network and provide only a local perspective from a router toward a destination.

*B. Traceroute AS graph*

There is currently another primary method [17] [19] for constructing AS graph by converting traceroute IP path using origin ASes for best-match prefixes for IP addresses. The CAIDA's topology measurement tool, Skitter [18], is a lightweight ICMP (Internet Control Message Protocol) traceroute tool designed to gather IP topology data. Skitter runs on more than 20 monitors around the globe and collects forward IP path and round trip times (RTTs) from more than one-half million destinations and captures the addresses of intermediate routers in the path.

Traceroute AS graph extracts interconnect information of ASes from the massive traceroute data collected by Skitter. Traceroute AS graph, using active probing methodology, is regarded to be more complete and reliable than the BGP AS graph [17] [19] [20].

*C. Barabási-Albert (BA) model*

By analyzing BGP AS graphs, Faloutsos *et al.* [1] discovered that the AS-level Internet topology is a scale-free network, which has a power-law degree distribution. The BA model [3] shows that a power-law degree distribution could arise from two generic mechanisms: 1) *growth*, where networks expand continuously by the addition of new nodes, and 2) *preferential attachment*, where new nodes are attached preferentially to *m* nodes that are already well connected. The probability $\Pi(i)$ that a new node will be connected to node *i* is

$$\Pi(i) = (k_i) / \sum_j (k_j) \qquad (1)$$

where $k_i$ is the degree of node $i$.

This degree-based model has generated great interests in various research areas from natural biology networks to manmade communication networks. A number of modifications of the BA model have been introduced ever since and the BA model has been applied to the research on error and attack tolerance of the Internet [21].

*D. Interactive Growth (IG) Model*

Recently we proposed the Interactive Growth (IG) model [12] [13], which features the joint growth of new nodes and new links. At each time-step, it has two inter-dependent operations that new nodes are added and connected to existing nodes (host nodes), and at the same time new links are added connecting the host nodes to other existing nodes (peer nodes). Host and peer nodes are chosen with the linear preference shown in equation (1).

The IG model resembles the actual evolution of Internet. New nodes bring new traffic load to its host nodes. This results in both the increase of traffic volume and the change of traffic pattern around host nodes and it triggers the addition of new links connecting host nodes to peer nodes in order to balance network traffic and optimize network performance.

## III. THE RICH-CLUB PHENOMENON

Power-law topologies have a small number of nodes having large numbers of links. We call these nodes 'rich nodes'. We reported [10] [11] that the AS-level Internet topology shows the rich-club phenomenon, in which rich nodes are very well connected to each other and rich nodes are preferentially connected to the other rich nodes. Rich nodes create a tight group, rich club (or the core tier), where if two club members do not share a link, it is very likely that they share a common node that can link them, such that the average hop distance within the club is between one and two. Since most nodes of the network are directly connected with one or more club members, the rich-club functions as a super traffic hub and it is crucial for network properties [13], such as routing efficiency, redundancy and robustness.

The rich-club phenomenon is parameterized by the rich-club connectivity and the node-node link distribution. The rich-club connectivity measures the interconnection between rich nodes. The node-node link distribution reflects to whom rich nodes are connected. The calculation of the two parameters is purely based on connectivity information without making any heuristic assumption on the interaction between network elements. And the complexity of calculating the two parameters is of order of $L$, number of links. Therefore the rich-club phenomenon can be used as a quantitatively simple metric to characterize the hierarchical structure of power-law topologies.

We used the rich-club to distinguish the topological structure of extended map of BGP AS graph, the BA model and the IG model. We found that the BA network does not have a rich-club at all that the best-connected nodes are not well connected to each other, instead, rich nodes are connected to nodes of all degrees with similar probabilities. However, the IG model accurately matches both the power-law degree distribution and the hierarchical structure of the BGP AS graph. We showed that this simply and dynamic model compares favourable with other available degree-based Internet topology generators.

TABLE I
TOPOLOGICAL PROPERTIES OF THE FOUR EXAMINED NETWORKS

| PROPERTIES | BGP AS GRAPH | TRACEROUTE AS GRAPH | IG MODEL | BA MODEL |
|---|---|---|---|---|
| $N$ | 11461 | 11122 | 11122 | 11122 |
| $L$ | 32730 | 30054 | 33349 | 33366 |
| $k_{average}$ | 5.7 | 5.4 | 6.0 | 6.0 |
| $k_{max}$ | 2432 | 2839 | 753 | 187 |
| $\varphi(r=1\%)$ | 31.7% | 28.0% | 34.4% | 4.6% |
| $l\,(r_i<5\%)$ | 87.4% | 91.0% | 76.1% | 45.4% |
| $l\,(r_i<5\%, r_j<5\%)$ | 27.3% | 19.5% | 22.3% | 4.4% |
| $P(1)$ | 28.9% | 26.1% | 26.0% | 0 |
| $P(2)$ | 40.3% | 37.9% | 33.3% | 0 |
| $P(3)$ | 11.6% | 14.0% | 10.6% | 39.4% |

$N$ – total number of nodes. $L$ – total number of links. $k_{average}$ – average degree. $k_{max}$ – maximum degree. $\varphi(r=1\%)$ – rich-club connectivity. $l(r_i<5\%)$ – percentage of links connecting *with* the 5% best connected nodes. $l(r_i<5\%, r_j<5\%)$ – percentage of links connecting *between* the 5% best connected nodes. $P(k)$ – degree distribution, percentage of nodes with degree $k$.

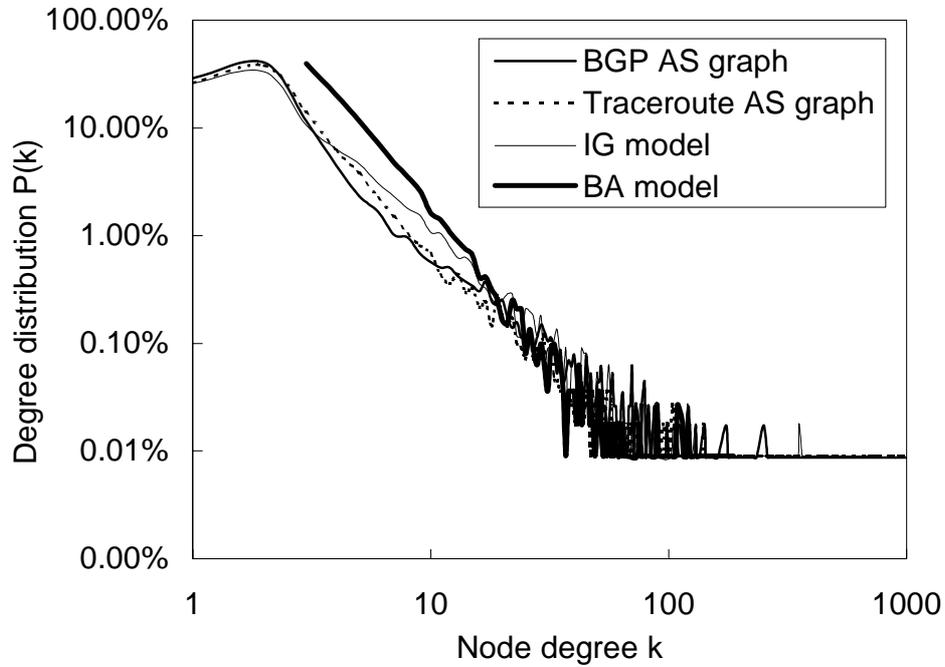

Figure 1. Degree distribution

To be prudent, in this paper we analyze and compare topological structures of Traceroute AS graph and BGP AS graph. Also we compare with two synthetic Internet topologies generated by the BA model and the IG model.

The traceroute AS graph was obtained from the CAIDA's Internet Topology Data Kit #0204 (available on request from the CAIDA [18]), which was measured on 1st April 2002. The extended map of BGP AS graph was measured on 26th May 2001 (dataset files are available on the web [16]). In this paper the IG model is grown as following, 1) with 40% probability, a new node is connected to one host node and the host node is connected to two peer nodes; and 2) with 60% probability, a new node is connected to two host nodes and one of the host nodes is connected to one peer node. Thus at each time-step, one new node and three new links are added. The BA model is grown with parameter $m=3$ and the minimum node degree is 3.

Table I shows that the four networks have similar numbers of nodes and links. Figure 1 and table I show that degree distributions of the two AS graphs are not strict power laws. For example, $P(k=1)$ is notably smaller than $P(k=2)$. The IG model closely matches this property. However the degree distribution of BA model is a strict power-law of $P(k) \sim k^{-3}$ [23].

*A. Rich-club connectivity*

When $n$ nodes are fully connected, they have the maximum possible number of links $n(n-1)/2$. The rich-club connectivity $\varphi(r)$ is defined as the ratio of the actual number of links over the maximum possible number of links between nodes with node rank less than $r$, where node rank $r$ is the rank of a node on a list sorted in a decreasing order of node degree and $r$ is normalized by the total number of nodes.

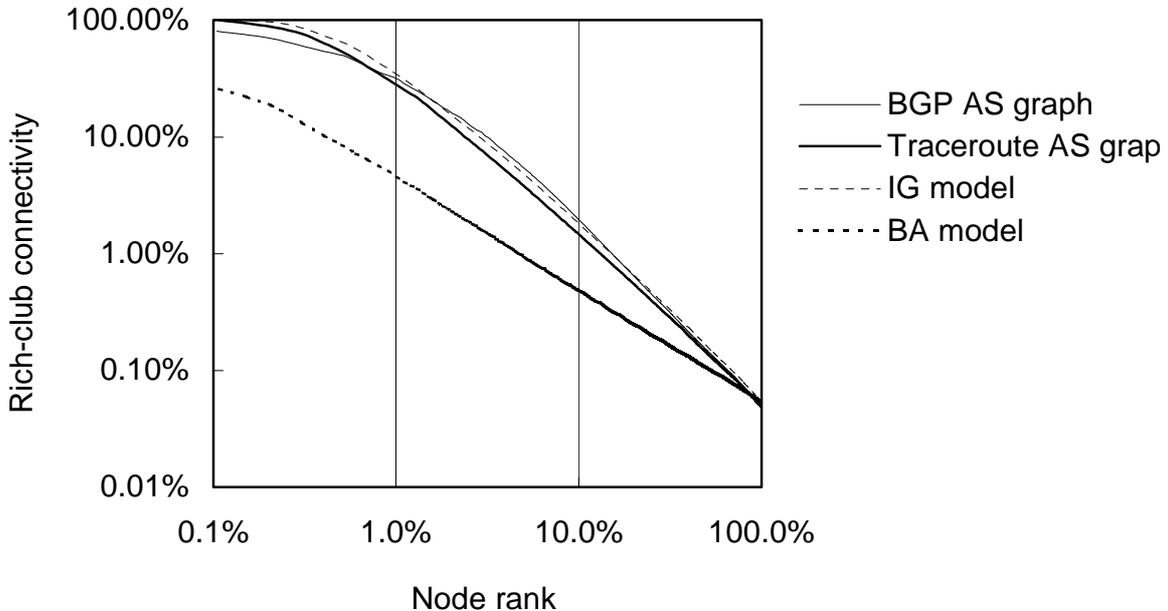

Figure 2. Rich-club connectivity $\varphi(r)$ against node rank $r$.

Figure 2 is a plot of the rich-club connectivity $\varphi(r)$ against node rank $r$ on a log-log scale. The plot shows that the rich-club connectivity of the two AS graphs and the IG model are similar and they are significantly larger than that of the BA model. Table I shows that $\varphi(1\%)$ of the two AS graphs and the IG model is around 28~34%, which means that the 1% best connected nodes have about 1/3 of the maximum possible number of links and indicates the three networks have a core tier in which rich nodes are densely connected to each other. Whereas the BA model does not show this property. Rich nodes of the BA model are not well connected between each other and $\varphi(1\%)$ is merely 4.6%.

*B. Node-node link distribution*

We define $l(r_i, r_j)$ as the percentage of links connecting nodes with node rank $r_i$ to nodes with node rank $r_j$, where node ranks are divided into 5% bins and $r_i \leq r_j$. Figure 3 illustrates the node-node link distribution by plotting $l(r_i, r_j)$ against corresponding node rank $r_i$ and $r_j$. In the two AS graphs and the IG model, 76-91% of the total links are connected *with* the top 5% rich nodes (black and dark gray columns) comparing with 46% in the BA model. This is also shown as *l($r_i$ <5%)* in Table I.

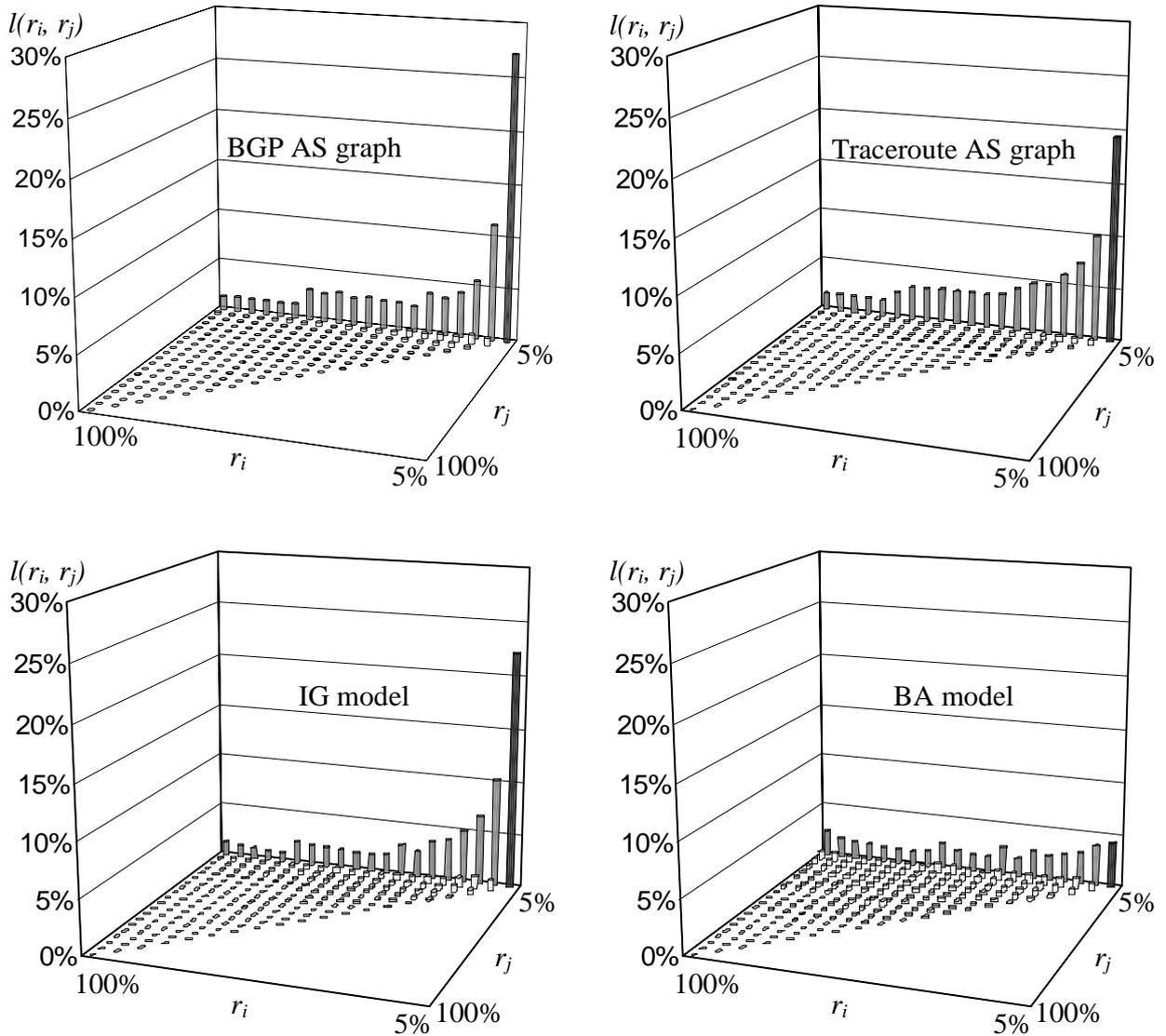

Figure 3. Node-node link distribution

The plot illustrates that the two AS graphs and the IG model show the rich-club phenomenon, in which rich nodes show a preference of connecting to other rich nodes that the percentage of links connecting *between* the 5% best connected nodes (black column, *l($r_i$<5% , $r_j$<5%)* in table I) is significantly larger than the percentages of links connecting these rich nodes to other lower degree nodes (dark gray columns). However the BA model does not show this phenomenon at all, in stead, rich nodes are connected to nodes of all degrees with similar probabilities.

## IV. CONCLUSIONS

Although the extended map of BGP AS graph and the Traceroute AS graph are constructed with different methodologies, both AS graphs show the rich-club phenomenon, in which a small number of best-connected nodes are very well connected to each other and rich nodes are preferentially connected to the other rich nodes. The calculation of rich-club phenomenon is purely based on connectivity information without making any heuristic assumption and it is a quantitatively simple way to differentiate topological structures of large-scale power-law topologies. By examining the rich-club phenomenon, we show that network structure of the BA model is fundamentally different from that of two AS graphs. However the IG model closely resembles the tier structure of the AS-level Internet topology.

A correct topology is important for modelling the Internet [13]. For example, the two AS graphs all have a densely inter-connected rich-club, which functions as a super traffic hub by providing a large amount of routing shortcuts. Therefore, topologies without the rich-club phenomenon could under-estimate the network routing efficiency and redundancy, and over-estimate the robustness under node failures, because the removal of only a few rich-club members can broken down the network integrity.